\documentclass{siamltex}
\usepackage[square,authoryear]{natbib}
\usepackage{amsmath,amssymb}
\usepackage[all]{xy}
\begin{document}

\title{Lagrangian Averaging for Compressible Fluids}
\author{H.S. Bhat\thanks{
Control and Dynamical Systems 107-81,
  California Institute of Technology,
  Pasadena, CA 91125. To whom correspondence should be addressed
  (bhat@cds.caltech.edu)} \and 
R.C. Fetecau\thanks{
Department of Mathematics,
Stanford University, Stanford, CA 94305-2125
(van@math.stanford.edu)}
\and
J.E. Marsden\thanks{
Control and Dynamical Systems 107-81,
  California Institute of Technology,
  Pasadena, CA 91125
  (marsden@cds.caltech.edu)}
\and 
K.Mohseni\thanks{
Department of Aerospace Engineering Sciences,
University of Colorado,
Boulder, CO 80309-0429
(mohseni@colorado.edu)}
\and 
M. West\thanks{Department of Aeronautics and Astronautics,
  Durand Building, Stanford University,
  Stanford, CA 94305-4035
  (westm@stanford.edu)}
}
\date{\today}
\maketitle

\begin{abstract}This paper extends the derivation of the Lagrangian averaged Euler
(LAE-$\alpha$) equations to the case of barotropic compressible flows.  The aim of Lagrangian averaging is to regularize the compressible Euler equations
by adding dispersion instead of artificial viscosity. Along the way, the
derivation of the isotropic and anisotropic LAE-$\alpha$ equations is
simplified and clarified. 

The derivation in this paper involves averaging over a tube of trajectories
$\eta^\epsilon$ centered around a given Lagrangian flow $\eta$.  
With this tube framework, the Lagrangian averaged Euler
(LAE-$\alpha$) equations are derived by following a simple procedure: start
with  a given action, Taylor expand
in terms of small-scale fluid fluctuations $\xi$, truncate, average, 
and then model those terms that are nonlinear functions of $\xi$. Closure of
the equations is provided through the use of \emph{flow rules}, which
prescribe the evolution of the fluctuations  along the mean flow.
\end{abstract}

\begin{keywords} averaged Lagrangians, inviscid compressible fluids.\end{keywords} 

\begin{AMS}37K99 37N10 76M30 76Nxx\end{AMS}

\pagestyle{myheadings}
\thispagestyle{plain}
\markboth{H. S. BHAT et.al}{LAGRANGIAN AVERAGING FOR COMPRESSIBLE FLUIDS}


\section{Introduction}

\paragraph{Historical Remarks}

The incompressible case will be discussed first. The Lagrangian averaged Euler
(LAE-$\alpha$) equations
for average incompressible ideal fluid motion first appeared in the context of
averaged fluid models in
\cite{HoMaRa1998a,HoMaRa1998b}.  Dissipation was added later to produce the
Lagrangian averaged Navier-Stokes (LANS-$\alpha$) equations, 
also known as the Navier-Stokes-$\alpha$ equations\footnote{Sometimes the
term ``viscous Camassa-Holm (VCH) equations'' (\cite{ChFoHoOlTiWy1998}) has
been used, but this terminology is a little unfortunate since the
$n$-dimensional version of the CH equations, also known as the EPDiff
equations, arise via Euler-Poincar\'e reduction of $H ^1$
geodesics on the group of all diffeomorphisms, and {\it not} the volume
preserving ones  (see
\cite{HoMa2004}).}.

Remarkably, the LAE-$\alpha$ equations are mathematically
identical to the inviscid second grade fluid equations introduced
in \cite{RiEr1955}, except for the fact that the parameter $\alpha$ is
interpreted differently in the two theories. In the case of LAE-$\alpha$ and LANS-$\alpha$, the parameter $\alpha$ is a spatial scale
below which rapid fluctuations are smoothed by linear and nonlinear
dispersion.

As in, for example, the work of \cite{Whitham1974} on nonlinear
waves, the distinctive feature of the Lagrangian averaging approach
is that averaging is carried
out at the level of the variational principle and not at the level of the
Euler or Navier-Stokes equations, which is the traditional averaging or
filtering approach used for both the Reynolds averaged Navier-Stokes (RANS) and the
large eddy simulation (LES) models. As such, the
variational procedure does not add any \emph{artificial} viscosity, a physical
reason to consider the LAE-$\alpha$ or LANS-$\alpha$ equations as good models
for incompressible turbulent flow. Moreover, it has been proven that the 
$\alpha$ models are computationally very attractive (see \cite{ChHoMaZh1999,MoKoShMa2003}). 

Although sharing the same general technique (use of averaging and asymptotic methods in the variational formulation), several alternative derivations of incompressible LAE-$\alpha$ equations exist in the literature. One of these derivations (see \cite{Holm1999}) uses the generalized Lagrangian mean (GLM) theory developed in \cite{AnMc1978}.

An alternative derivation of the incompressible LAE-$\alpha$ and
LANS-$\alpha$ equations was given in \cite{MaSh2002} by using an
ensemble average over the set of solutions of the Euler equations with
initial data in a phase-space ball of radius $\alpha$, while treating the
dissipative term via stochastic variations. The derivation also uses a
turbulence closure that is based on the Lagrangian fluctuations, namely a
generalization of the frozen turbulence hypothesis of Taylor (see
\cite{Taylor1938}).

Rigorous analysis aimed at proving global well-posedness and regularity of
the three-dimensional isotropic and anisotropic LANS-$\alpha$ equations can
be found in, for example, \cite{FoHoTi2002,MaSh2001,MaSh2002}. However,
global existence for the inviscid three-dimensional Lagrangian averaged Euler
(LAE-$\alpha$) remains an open problem.

From a computational viewpoint, numerical simulations of the $\alpha$ models
(see \cite{ChHoMaZh1999,MoKoShMa2003}) show that the LANS-$\alpha$ equations
give comparable computational savings as LES models for forced and decaying
turbulent flows in periodic domains. For wall-bounded flows, it is expected
that either the anisotropic model or a model with varying $\alpha$ needs to be
used; the computational efficacy of these methods on such flows remains to be
demonstrated.

As far as the compressible case is concerned, the only reference we know of
is \cite{Holm2002}. We shall discuss the relation between the work in this
reference and the present paper below.

 We refer the interested reader to \cite{MaSh2001,MaSh2002} 
 for a more detailed history of the PDE analysis for
LAE-$\alpha$ and LANS-$\alpha$ equations and to \cite{MoKoShMa2003} for a
survey and further references about the numerical aspects of these models.

\paragraph{Motivation}
In compressible flows there are two major problems at higher wave
numbers, or small scales, that require special attention. These are (a)
turbulence for high Reynolds number flows (common with incompressible
flows) and (b) strong shocks. In both cases the challenge lies in the
appropriate representation of small scale effects. For turbulence, the
energy cascade to smaller scales can be balanced by viscous
dissipation, resulting in the viscous regularization of the Euler
equations.

Historically, viscous dissipation has been used to  regularize shock
discontinuities. This includes adding to the Euler equation {\it non-physical
and artificial viscous terms} and Fourier's law for heat transfer in the shock
region (see e.g., \cite{LiRo1957, Shapiro1953}). This way, the steepening
effect of the nonlinear convective term is balanced by dissipation. We
believe that Lagrangian averaging is a reasonable alternative way to
regularize shock waves.  The net effect of Lagrangian averaging  is to add
dispersion instead of dissipation to the Euler equations; that is, one adds
terms that redistribute energy in a nonlinear fashion.  In other, rather
different situations, the technique of balancing a nonlinear convective term
by dispersive mechanisms was used  by \cite{LaLe1983} for the
KdV equation and by
\cite{Kawahara1970, KaKa1970} for plasma flows.

The competition between nonlinearity and dispersion has of course resulted in
remarkable discoveries, the most famous being solitons, localized waves
that collide elastically, suffering only a shift in phase.  The robustness of
solitons in overcoming strong perturbations is largely due to a balance
between nonlinearity and linear dispersion. Note that in Lagrangian
averaging, the energy redistribution mechanism that is introduced is nonlinear
and might yield other interesting features that warrant further investigation.

Another feature of the Compressible Lagrangian Averaged
Navier-Stokes-$\alpha$ equations (or CLANS-$\alpha$
equations)  is that in turbulent flows with shocks, the effect of
shocks and turbulence are simultaneously modeled by the same technique,
namely the Lagrangian averaging method. 

\paragraph{Issues Addressed in This Paper}
In this paper we apply the averaged Lagrangian methodology to derive the
isotropic and anisotropic averaged models for compressible Euler equations.

One goal of this paper is to present a clear derivation of the averaged
equations. We are particularly interested in separating the two issues of
averaging and modeling. In the derivation, a new ensemble averaging 
technique is proposed and investigated. Instead of taking clouds of initial conditions, as in \cite{MaSh2002}, we average over a tube of trajectories
$\eta^{\epsilon}$ centered around a given Lagrangian flow $\eta$. The tube is
constructed by specifying the Lagrangian fluctuations $\xi^{\epsilon}=
\eta^{\epsilon} \circ \eta^{-1}$ at $t=0$ and providing a {\it flow rule}
which evolves them to all later times. The choice of flow rule is a precise
modeling assumption which brings about closure of the system.

For the incompressible case we assume that fluctuations are Lie advected by
the mean flow (or frozen into the mean flow as divergence-free vector fields),
and we obtain both the isotropic and the anisotropic versions of the
LAE-$\alpha$ equations. The advection hypothesis is the natural extension to
vector fields of the classical Frozen Turbulence Hypothesis of G.I. Taylor
(see \cite{Taylor1938}) stated for scalar fluctuations.

The second goal of this work is to extend the derivation to barotropic
compressible flows. This problem has already been considered by Holm (see
\cite{Holm2002}) in the context of generalized Lagrangian mean (GLM) motion.
In this work, an alpha model appears as a GLM fluid theory with an
appropriate Taylor hypothesis closure. However, even though \cite{Holm2002}
enumerates several frozen-in closure hypotheses, the averaged equations are
derived only for the case when the fluctuations are parallel transported by
the mean flow. In our work we will consider a more general advection
hypothesis to study the compressible anisotropic case. In addition, a
physically based new flow rule is introduced to deal with the
isotropic case.

The averaging technique consists of expanding the original Lagrangian with 
respect to a perturbation parameter $\epsilon$, truncating 
the expansion to $O(\epsilon^2)$ terms, and then taking the average. 
It turns out that the averaged
compressible Lagrangian depends on the Lagrangian fluctuations $\xi'$ only
through three tensor quantities which are quadratic in $\xi'$. In the
terminology of \cite{Holm2002} these tensors represent the second-order
statistics of the Lagrangian fluctuations.  Evolution equations for
these tensors are derived from a core modeling assumption: a prescribed
\emph{flow rule} for the time-evolution of the fluctuations $\xi'$.  The flow
rule gives us closure, allowing us to apply Hamilton's principle to the
averaged Lagrangian and thereby derive an equation for the mean velocity $u$.

The organization of the rest of the paper is as follows. In
\S\ref{averaging-general} we describe a general procedure for Lagrangian
ensemble averaging. This procedure is then applied to the action for
incompressible fluids in \S\ref{averaging-incomp} to demonstrate our
derivation technique.  The general procedure is applied again in \S\ref{averaging-Euler}, this time to the more complex case of barotropic
compressible fluids. \S\ref{flow-rules} is devoted to modeling issues; here
the strategy of modeling the evolution of Lagrangian fluctuations $\xi'$
using flow rules is discussed in detail. In \S\ref{averaged-pdes} we derive the
averaged equations for incompressible and compressible models in both
isotropic and anisotropic versions. The Appendix provides technical details
about the fluctuation calculus used throughout the paper.

\paragraph{Main Results}
The main result of this paper is the derivation of compressible
Lagrangian-averaged Euler equations with
\begin{itemize}
\item anisotropic modeling of fluid 
fluctuations---see equations (\ref{eqn:avg_advection_pde_short}).
\item isotropic modeling of fluid 
fluctuations---see equations (\ref{eqn:avg_comp_pde_isotropic}).
\end{itemize}
In addition, we provide an improved derivation of the {\it incompressible}
isotropic and anisotropic LAE-$\alpha$ equations. 


\section{General Lagrangian Averaging}
\label{averaging-general}

A mathematical setting for a certain class of compressible
fluid flow problems will first be given. After describing the general
procedure for Lagrangian averaging, the specific case of the Euler 
action for fluids will be considered.

Let $M$ be an open subset of $\mathbb{R}^N$, representing the containing 
space of a fluid. Suppose we are given a Lagrangian for a compressible fluid, $L(\psi, \dot{\psi}, \mu_0)$, where $\psi \in \text{Diff}(M)$ the space of
diffeomorphisms of $M$, $(\psi, \dot{\psi}) \in T
\text{Diff}(M)$ and $\mu_0 \in \Lambda^N(M)$, the
space of $N$-forms on $M$. Fix a time interval $[0,T]$ and let
$\mathcal{C}(\text{Diff}(M))$ be the path space of smooth maps from
$[0,T]$ into $\text{Diff}(M)$. Then the action 
$S : \mathcal{C}(\text{Diff}(M)) \times \Lambda^N(M) \to \mathbb{R}$ is 
$$
S(\eta, \mu_0) = \int_{0}^{T} L(\eta(t),\dot{\eta}(t),\mu_0) \, dt.
$$
We seek an averaged action $S^{\alpha}(\eta, \mu_0)$, where $\alpha$ is
a length scale characterizing the coarseness of the average.  Taking $\eta$
and $\mu_0$ as given, we shall describe how to compute
$S^{\alpha}(\eta,\mu_0)$.

\paragraph{Remark} It is important to emphasize that for both $S$ and
$S^{\alpha}$, $\eta$ is merely a test curve. It is \emph{not} an extremal of
the action $S$.  We are trying to average the action $S$ itself, not any fluid
dynamical PDE or the  solutions of such a PDE.  Our final product
$S^{\alpha}$ should not depend at all on an initial choice of the test curve
$\eta$.

\paragraph{Tube Initialization}
The first step is to take $\xi^{\epsilon}(x,t)$ to be a family of diffeomorphisms about the identity.
That is,
\begin{align*}
\text{for each } \epsilon &\geq 0, \quad \xi^{\epsilon}(\cdot,t) \in \text{Diff}(M) \text{ for all } t, \text{ and} \\
\text{at } \epsilon &= 0, \quad \xi^{\epsilon}(x, t) = x \text{ for all }x, t.
\end{align*}
Define the vector fields $\xi'$ and $\xi''$ via
$$
\xi' = \left. \frac{\partial}{\partial \epsilon} \right|_{\epsilon = 0} \xi^{\epsilon} \ \text{ and } \
\xi'' = \left. \frac{\partial^2}{\partial \epsilon^2} \right|_{\epsilon = 0} \xi^{\epsilon}. 
$$
Use $\xi^{\epsilon}$ to construct a tube of material deformation maps
that are close to $\eta$ by letting
$\eta^{\epsilon}(X, t) = \xi^{\epsilon}(\eta(X, t), t)$, or, written more
compactly,
\begin{equation}
\label{eqn:eta_mod}
\eta^{\epsilon} = \xi^{\epsilon} \circ \eta.
\end{equation}
Here, $X$ is a material point in the reference configuration.
Define the spatial velocity by
$u^{\epsilon}(x,t) = \dot{\eta}^{\epsilon}((\eta^{\epsilon})^{-1}(x, t), t)$,
where $\eta^{\epsilon}$ is a given material deformation map.  Compactly written,
this reads
\begin{equation}
\label{eqn:u_mod}
u^{\epsilon} = \dot{\eta}^{\epsilon} \circ (\eta^{\epsilon})^{-1}.
\end{equation}
The map $u^{\epsilon}$ is a time-dependent vector field on $M$, i.e.
for each $\epsilon \geq 0$, and for all $t$, 
$u^{\epsilon}(\cdot, t) \in \mathfrak{X}(M)$.

\paragraph{Averaging} The existence of an averaging
operation $\left\langle \, \cdot \, \right\rangle$ will now be postulated.
The properties this operation is required to satisfy and an example of such
an operation will be given shortly.
 
\paragraph{Relationship Between $u^\epsilon$ and $u$}
It is desirable to have the fluctuations $\xi^{\epsilon}$ centered, on
average, about the identity: $\langle \xi^{\epsilon} (x,t)\rangle = x \text{
for all positions } x \text{ at all times } t$.  What is actually needed is
that for $n \geq 1$,
\begin{equation}
\label{eqn:xi-centered-meaning}
\Biggl \langle \left. \frac{\partial^n \xi^\epsilon}{\partial \epsilon^n} \right|_{\epsilon = 0} \Biggr \rangle = 0.
\end{equation}
In other words, the $n$-th order fluid fluctuation vector fields should all have mean zero.   Restricting the map to be centered about the identity means simply that the
average will not be skewed in an arbitrary direction.
From (\ref{eqn:u_mod}) and (\ref{eqn:xi-centered-meaning}) one can derive
\begin{equation}
\label{eqn:u-centered}
\langle u^{\epsilon} \circ \xi^{\epsilon} (x,t) \rangle  = u(x,t).
\end{equation}
Equation (\ref{eqn:u-centered}) shows in which sense the average of $u^{\epsilon}$ is $u$ in a Lagrangian mean theory defined by $\langle \eta^{\epsilon} (\cdot,t) \rangle = \eta(\cdot,t)$. This equation is closely connected with the generalized Lagrangian-mean description of \cite{AnMc1978}, where the Lagrangian mean velocity $\bar{u}^L$ and the fluctuating Eulerian velocity $u^{\xi}$ are related in a similar way. 

\paragraph{Density}  For the non-averaged Lagrangian $L$, $\mu_0$ is a parameter in the sense
of Lagrangian semidirect product theory; see \cite{MaRaWe1984, HoMaRa1998b}.
The physical interpretation of $\mu_0$ is as follows.
Since $\mu_0$ is an $N$-form on $M$, it can be written as
$$
\mu_0 = \rho_0 \, dx^1 \wedge \cdots \wedge dx^N
$$
where $\rho_0$ is a smooth function on $M$.
Now $\rho_0(X)$ is the density of the fluid at the material
point $X$ in the reference configuration.  This is in contrast
to the spatial density $\rho^{\epsilon}(x,t)$, which gives us the density 
of the fluid at the spatial point $x$ at time $t$.  Defining
\begin{equation}
\label{eqn:mu_defs}
\mu^\epsilon = \rho^\epsilon \, dx^1 \wedge \cdots \wedge dx^N,
\end{equation}
one has the relationship
\begin{equation}
\label{eqn:rho_mod}
(\eta^{\epsilon})_* \mu_0 = \mu^\epsilon.
\end{equation}

\paragraph{Fluctuation Calculus}
Because $u^\epsilon$ and $\rho^\epsilon$
will be expanded,  the $\epsilon$-derivatives of
$u^{\epsilon}$ and
$\rho^{\epsilon}$ need to be calculated. First,  define
\begin{equation}
\label{eqn:u_prime_defs}
u' = \left. \frac{\partial}{\partial \epsilon} 
 \right|_{\epsilon = 0} u^{\epsilon},
\qquad \text{and} \qquad
u'' = \left. \frac{\partial^2}{\partial \epsilon^2} 
         \right|_{\epsilon = 0} u^{\epsilon}.
\end{equation}
By differentiating (\ref{eqn:u_mod}), one finds expressions for $u'$ and $u''$ in 
terms of $u$, $\xi'$, and $\xi''$.  The calculations can be
performed  intrinsically using Lie derivative formulae --- the results, as
found in \cite{MaSh2002}, are
\begin{subequations}
\label{eqn:u_primes}
\begin{align}
\label{eqn:u_prime}
u' &= \partial_t \xi' + [u, \xi'], \\
\label{eqn:u_prime_prime}
u'' &= \partial_t \xi'' + [u, \xi''] - 2 \nabla u' \cdot \xi' - 
\nabla \nabla u (\xi', \xi').
\end{align}
\end{subequations}
In these formulas, the bracket $[x, y] = \pounds_{x} y$ is the standard 
Jacobi-Lie bracket of vector fields on $M$ (see, for example,
\cite{AbMaRa1988}).  
Next, define
\begin{equation}
\label{eqn:rho_prime_defs}
\rho' = \left. \frac{\partial}{\partial \epsilon} \right|_{\epsilon = 0} \rho^{\epsilon},
\qquad \text{ and } \qquad
\rho'' = \left. \frac{\partial^2}{\partial \epsilon^2} \right|_{\epsilon = 0} \rho^{\epsilon}.
\end{equation}
One obtains expressions for $\rho'$ and $\rho''$ in terms of $\rho$, $\xi'$, 
and $\xi''$ by differentiating (\ref{eqn:rho_mod}) (see the appendix for
the detailed calculations). The results are:
\begin{subequations}
\label{eqn:rho_primes}
\begin{align}
\label{eqn:rho_prime}
\rho' &= -\operatorname{div}(\rho \xi'), \\
\label{eqn:rho_prime_prime}
\rho'' &= \operatorname{div}(\operatorname{div}(\rho \xi' \otimes \xi')) - \operatorname{div}(\rho \xi'').
\end{align}
\end{subequations}

\paragraph{Averaging Operation}
In the above development, an averaging operation has been 
implicitly used. The properties it is required to satisfy will now be
spelled out.  Let $\mathcal{F}(Y)$ mean the space of smooth, real-valued
functions on a manifold $Y$. If $Y$ is infinite dimensional, then smoothness
is understood in the sense of infinite dimensional calculus with respect to,
for example, suitable Sobolev topologies. These infinite
dimensional technicalities will not be required in any detail in this paper,
and so may be treated formally.

As before, the set $M$ is the containing space of the fluid and
$\alpha$ is a small positive number. Let $\mathfrak{X}$ be an
appropriately chosen space of fields, designed to model ``fluid
fluctuations,'' on $M$, and consider the space $Y = [0, \alpha] \times
\mathfrak{X}$.  Assume that there is an averaging operation
\[
 \left\langle \, \cdot \, \right\rangle : \mathcal{F} (Y) \rightarrow
\mathcal{F} (M)
\]
satisfying the following properties for $f,g \in \mathcal{F}(Y)$, $a,b
\in\mathbb{R}$, $ \psi \in \mathcal{F}([0, \alpha] ) $, and $h \in \mathcal{F}
(\mathfrak{X})$,
\begin{align}
\label{eqn:axiom1}
\text{Linearity: }&\langle a f + b g \rangle = a \langle f
\rangle + b \langle g \rangle, \\
\label{eqn:axiom2}
\text{Independence: }& 
\langle \psi h \rangle  = \frac{1}{\alpha} \left(
\int_{0}^{\alpha} \psi(\epsilon) \, d \epsilon
\right) \langle h \rangle, \\
\label{eqn:axiom3}
\text{Commutativity: }&\Biggl \langle \int f \, dx \Biggr \rangle = \int
\langle f \rangle \, dx, \\
\label{eqn:axiom4} &\langle \partial f \rangle = \partial \langle f \rangle,
\text{ where }
\partial = \partial_t \text{ or } \partial = \partial_{x^i}.
\end{align} 

Here, $\psi h \in \mathcal{F} (Y)$ is defined as the pointwise
product. Note that if $\psi$ is a constant, then the first and second
requirements are compatible. 

For compressible flow, the space of fluid fluctuations is
$\mathfrak{X} = \mathfrak{X}(M)$.  For  incompressible flow, the space of
divergence-free vector fields is used instead, i.e. $\mathfrak{X} = \mathfrak{X}_{\text{div}}(M)$.  In general, $\mathfrak{X} =
T_{\text{Identity}}X$, where $X$ is  the space to which the tube maps
$\xi^\epsilon$ belong.

\paragraph{Example}
Let $\mu$ be a probability measure on the unit sphere $S$ in
$\mathfrak{X}(M)$, and define the \emph{average} of a (vector-valued)
function $f(\epsilon, w)$ on $[0, \alpha] \times S$ by
$$
\langle f \rangle := \frac{1}{\alpha} \int_{0}^{\alpha} \int_{S} 
f(\epsilon, w) \, d\mu(w) \, d \epsilon.
$$

One checks formally that this is an example of an averaging operation
that satisfies the desired properties.


\section{Incompressible Flow Revisited}
\label{averaging-incomp}

Before applying the averaging technique to the case of compressible flow,
we shall first derive averaged equations for incompressible flow, equations
which have already been derived in the literature.  The presentation given
here has the advantage of being easily generalized to compressible flows. This
advantage stems from the careful use and interpretation of modeling 
assumptions on the fluctuations $\xi'$ --- only intuitive
assumptions are required regarding the mean behavior of the fluctuations as
well as a first-order Taylor hypothesis.  Furthermore, great care has been taken
to separate the algebraic issues involved with the averaging procedure
from the modeling issues.

In the incompressible case, fluid fluctuations are modeled using the 
\emph{volume-preserving} diffeomorphism group on $M$ which is denoted
by $\text{Diff}_{\text{vol}}(M)$.  Therefore, the tube construction
from the previous section now reads:
let $\xi^{\epsilon}(x,t)$ be a family of volume-preserving
diffeomorphisms about the identity.
That is,
\begin{align*}
\text{for each } \epsilon &\geq 0, \quad \xi^{\epsilon}(\cdot,t) \in \text{Diff}_{\text{vol}}(M) \text{ for all } t, \text{ and} \\
\text{at } \epsilon &= 0, \quad \xi^{\epsilon}(x, t) = x \text{ for all }x, t.
\end{align*}
This forces $\xi'(\cdot, t)$ to be a divergence-free vector field for all $t$.

\paragraph{Averaged Lagrangian for Incompressible Fluids}  Let us start with the
standard Lagrangian
\begin{equation}
\label{eqn:unavg_l_incomp}
l(u^\epsilon) = \int_{M} \frac{1}{2} \| u^\epsilon \|^2 \ dx,
\end{equation}
and expand $u^\epsilon$ in a Taylor series about $u$:
\begin{equation}
\label{eqn:u_expan}
u^{\epsilon} = u + \epsilon u' + \frac{1}{2} \epsilon^2 u'' + \mathcal{O}(\epsilon^3). 
\end{equation}
Substituting this expansion into (\ref{eqn:unavg_l_incomp}) gives
\begin{equation}
\label{eqn:uli_expan1}
l(u^\epsilon) = \int_{M} \frac{1}{2} \| u^2 \| + \epsilon u \cdot u' 
+ \frac{\epsilon^2}{2} \left( \| u' \|^2 + u'' \cdot u \right) + 
\mathcal{O}(\epsilon^3) \ dx.
\end{equation}
Let $\hat{l}(u^\epsilon)$ be the truncation of $l$ to terms of order 
less than $\epsilon^3$.  Using formulas (\ref{eqn:u_primes}), 
$u'$ and $u''$ can be rewritten in terms of $u$, $\xi'$, and $\xi''$.  We do
this in order to write $\hat{l}$ as a function only of $u$, $\xi'$, and $\xi''$.
Making the substitutions and rewriting in coordinates,
\begin{align}
\hat{l}(u^\epsilon) &= \int_{M} \frac{1}{2} u^i u^i 
+ \epsilon \Bigl( u^i (\partial_t {\xi'}^i)
+ u^i u^j {\xi'}^{i}_{,j} - u^i {\xi'}^j u^{i}_{,j} \Bigr)
+ \frac{\epsilon^2}{2} \Bigl(
(\partial_t {\xi'}^{i}) (\partial_t {\xi'}^{i})	\nonumber \\
& \quad + 2 (\partial_t {\xi'}^{i}) {\xi'}^{i}_{,k} u^k 	
- 2 (\partial_t {\xi'}^{i}) u^{i}_{,k} {\xi'}^{k}
+ {\xi'}^{i}_{,j} u^j {\xi'}^{i}_{,k} u^{k}
- {\xi'}^{i}_{,j} u^j u^i_{,k} {\xi'}^k		\nonumber \\
& \quad - u^i_{,j} {\xi'}^j {\xi'}^{i}_{,k} u^k 
+ u^i_{,j} {\xi'}^j u^i_{,k} {\xi'}^{k}
- 2 (\partial_t {\xi'}^{i}_{,j}) {\xi'}^{j} u^i
- 2 {\xi'}^{i}_{,jk} u^k {\xi'}^{j} u^i		\nonumber \\
& \quad - 2 {\xi'}^{i}_{,k} u^k_{,j} {\xi'}^{j} u^i
+ 2 u^{i}_{,kj} {\xi'}^{k} {\xi'}^{j} u^i
+ 2 u^i_{,k} {\xi'}^{k}_{,j} {\xi'}^j u^i
- u^i_{,jk} {\xi'}^{j} {\xi'}^{k} u^i
\Bigr)						\nonumber \\
\label{eqn:uli_expan_coords}
& \quad + \frac{\epsilon^2}{2} \Bigl(
(\partial_t {\xi''}^i) u^i 
+ u^j {\xi''}^{i}_{,j} u^i
- {\xi''}^{j} u^{i}_{,j} u^i
\Bigr) \ dx,
\end{align}
where the notation $u^i_{,j}$ means
$\partial u^i / \partial x^j$.
Throughout this paper, there is an implied sum over repeated indices.
The averaged Lagrangian for 
incompressible flow is now simply $l^{\alpha}_{\text{in}} 
= \langle \hat{l} \rangle$.

\paragraph{Zero-Mean Fluctuations} Before undertaking this computation, 
recall from \S\ref{averaging-general} that the fluctuation
diffeomorphism maps $\xi^\epsilon$ are required to have as their average the
identity map.  This statistical assumption regarding the behavior of the
fluctuations is the first modeling assumption:
\begin{equation}
\label{eqn:centered_fluc_model}
\langle \xi' \rangle = 0 \ \ \text{ and } \ \ \langle \xi'' \rangle = 0.
\end{equation}
This point would not be worth belaboring except that, when 
combined with
the properties of our averaging operation (\ref{eqn:axiom1}-\ref{eqn:axiom4}), assumption (\ref{eqn:centered_fluc_model}) forces \emph{all} linear functions of $\xi'$, $\xi''$, and their derivatives to also have zero mean.  Applying this fact to (\ref{eqn:uli_expan_coords}) causes the entire $\mathcal{O}(\epsilon)$ group and
the second $\mathcal{O}(\epsilon^2)$ group (i.e. the last line of (\ref{eqn:uli_expan_coords})) to vanish inside the average.

We continue analyzing (\ref{eqn:uli_expan_coords}): the only remaining terms are $(1/2) u^i u^i$ and the first $\mathcal{O}(\epsilon^2)$ group.  Within this $\mathcal{O}(\epsilon^2)$ group, we integrate certain terms by parts and notice that all terms involving time-derivatives of $\xi'$ group together:
\begin{multline}
\label{eqn:mat_deriv_parts}
(\partial_t {\xi'}^{i}) (\partial_t {\xi'}^{i})
+ 2 (\partial_t {\xi'}^{i}) {\xi'}^{i}_{,k} u^k
+ {\xi'}^{i}_{,j} u^j {\xi'}^{i}_{,k} u^{k} \\
=
\left( (\partial_t {\xi'}^{i}) + {\xi'}^{i}_{,j} u^j \right)
\left( (\partial_t {\xi'}^{i}) + {\xi'}^{i}_{,k} u^k \right)
= \left\| \frac{D \xi'}{D t} \right\|^2,
\end{multline}
where $D/Dt$ is the material derivative:
\begin{equation}
\label{eqn:mat_deriv_defn}
\frac{D}{Dt} = \left( \partial_t + u \cdot \nabla \right).
\end{equation}
We then simplify the remaining 
non-time-derivative terms from (\ref{eqn:uli_expan_coords}),
integrating by parts to remove second-order spatial derivatives.
The final expression for the averaged incompressible Lagrangian is
\begin{equation}
\label{eqn:avg_incomp_aniso_l_expr}
l^{\alpha}_{\text{in}}(u) = \int_{M} \left\{\frac{1}{2}
\| u \|^2 +\frac{\alpha^2}{2} 
\left[\left \langle \left \| \frac{D \xi'}{Dt} \right \|^2
\right \rangle - \frac{1}{2} \left \langle 
\operatorname{tr} (\nabla \xi' \cdot \nabla \xi') \right \rangle \|u\| ^2
\right] \right\}dx.
\end{equation}

\paragraph{Modeling of $\xi'$} Immediate application of Hamilton's
principle to (\ref{eqn:avg_incomp_aniso_l_expr}) does not yield a closed
system of equations.  Namely, we have
initial ($t = 0$) data for $\xi'$ but no way to compute this vector field for
$t > 0$.  Our approach in what follows will be to write down, based on
physical considerations, an evolution law, or {\em flow rule}, for $\xi'$. 

A flow rule consists of a prescribed choice of $\phi$ in the following 
evolution equation for $\xi'$:
\begin{equation}
\label{eqn:flow_rule}
\frac{D \xi'}{Dt} = \phi(u, \rho, \xi'). 
\end{equation}
Given a choice of $\xi'$ at $t = 0$, 
this equation will uniquely determine $\xi'$ for $t > 0$.
Let us assume we have a \emph{linear flow rule},
\begin{equation}
\label{eqn:flow_rule_linear}
\frac{D {\xi'}^i}{Dt} = \Omega^{ij} {\xi'}^j,
\end{equation}
where $\Omega^{ij}$ is allowed to depend on 
$u$ and $\rho$ but not on $\xi^\epsilon$ or its derivatives.
The caveat here is that our choice of $\Omega$ must be compatible with
incompressibility; in particular, $\operatorname{div} \xi' = 0$
at $t = 0$, and $\Omega$ must be chosen such that $\xi'$ remains divergence
free as it evolves.  At this stage, one might raise the issue of the tube $\xi^\epsilon$ and request a concrete description of the whole object.
Such a description is unnecessary; in order to close 
the system of evolution equations
resulting from (\ref{eqn:avg_incomp_aniso_l_expr}), 
we need only describe the evolution of the first-order fluctuation field
$\xi'$. Now defining the \emph{Lagrangian covariance tensor}
\begin{equation}
\label{eqn:cov_tens}
F = \langle \xi' \otimes \xi' \rangle
\end{equation}
and using the linear flow rule (\ref{eqn:flow_rule_linear}), 
the Lagrangian (\ref{eqn:avg_incomp_aniso_l_expr}) can be rewritten as
\begin{equation}
\label{eqn:avg_incomp_l_expr_2}
l^{\alpha}_{\text{in}}(u) = \int_{M} \left\{ \frac{1}{2} u^i u^i
+ \frac{\alpha^2}{2} \left[ \Omega^{ij} \Omega^{ik} F^{jk} - \frac{1}{2}
F^{ij}_{,ij} u^k u^k \right] \right\} \, dx.
\end{equation}
Here we have used the fact that $\xi'$ must be divergence-free.

\paragraph{Advection Flow Rule} 
The first flow rule we shall consider results from setting
$\Omega^{ij} = u^i_{,j}$:
\begin{equation}
\label{incompFL1}
\frac{D {\xi'}^i}{D t} = u^i_{,j} {\xi'}^j.
\end{equation}
Using the definition of the material derivative, it is trivial to see that
this flow rule is equivalent to Lie advection of $\xi'$: $\partial_t \xi' =
-\pounds_{u} \xi'.$ This advection hypothesis is the vector field
analogue of the classical Frozen Turbulence Hypothesis of G.I. Taylor
introduced in \cite{Taylor1938}. This hypothesis is widely used in the turbulence
community (see \cite{Cocke1969} for instance for usage of this hypothesis
even in the sense of Lie advection of vector fields). More recently, this
generalized version of Taylor hypothesis has been used to achieve turbulence
closure in the derivation of incompressible LAE-$\alpha$ equations (see
\cite{MaSh2001,MaSh2002}) or in the work of Holm (see \cite{Holm2002}) on
averaged compressible models using the generalized Lagrangian mean (GLM)
theory.   

The advection flow rule (\ref{incompFL1}) is perhaps the most obvious choice
for $\Omega$ that is compatible with incompressibility.  Note that if
$\operatorname{div} \xi' = 0$ at $t = 0$, then differentiating
(\ref{incompFL1}) with respect to $x^i$ yields
$$
\partial_t \left( \operatorname{div} \xi' \right) = u^{i}_{,j} {\xi'}^j_{,i} - {\xi'}^i_{,j} u^{j}_{,i} = 0.
$$
Therefore, $\operatorname{div} \xi' = 0$ for all $t > 0$.  Using this flow rule, 
both anisotropic and isotropic models shall be developed.  For incompressible
flow, no other flow rules will be considered.

\paragraph{Incompressible, Anisotropic, Inhomogeneous Flow} In this case, the flow rule is used to derive an evolution
equation for the covariance tensor $F$.  Time-differentiating
$F^{ij} = \langle {\xi'}^i {\xi'}^j \rangle$ and using (\ref{incompFL1})
yields the Lie advection equation $\partial_t F = - \pounds_{u} F$. 
Equipped with an evolution equation for $F$, we can apply Hamilton's
principle to  (\ref{eqn:avg_incomp_l_expr_2}) and derive a closed system with
unknowns $u$, the average velocity, and $F$, the covariance tensor.

Carrying this out, one finds that the anisotropic LAE-$\alpha$ equations are
given by the following coupled system of equations for $u$ and $F$:
\begin{subequations}
\label{eqn:LAE_in_anis}
\begin{align}
\label{eqn:LAE_in_anis_1}
\partial_t(1-\alpha^2 C)u + (u \cdot \nabla)(1-\alpha^2 C)u &
= - \operatorname{grad} p, \\
\label{eqn:LAE_in_anis_2}
\operatorname{div} u &= 0,\\
\label{eqn:LAE_in_anis_3}
\partial_t F + \nabla F \cdot u - F \cdot \nabla u - \nabla u^T \cdot F &= 0,
\end{align}
\end{subequations} 
where $p$ is the fluid pressure, and the operator $C$ is defined by
\begin{equation}
\label{eqn:op_C}
Cu = \operatorname{div}[\nabla u \cdot F].
\end{equation}
When $\alpha = 0$, the system (\ref{eqn:LAE_in_anis_1}-\ref{eqn:LAE_in_anis_2}) 
reduces to the incompressible Euler equation.

\paragraph{Note} 
Start with the generic incompressible averaged Lagrangian (\ref{eqn:avg_incomp_l_expr_2}) and substitute the advection flow rule 
(\ref{incompFL1}).  Now integrate the last term by parts and
use $\operatorname{div} \xi' = 0$.  The result is
\begin{equation}
\label{eqn:avg_incomp_aniso_l_expr_1}
l^{\alpha}_{\text{in}}(u) = \int_{M} 
\left\{\frac{1}{2} \left \|u \right \|^2 - \frac{\alpha^2}{2} u \cdot \left[
\nabla \nabla u :F \right] \right\}dx,
\end{equation}
which is exactly the Lagrangian used in \cite{MaSh2002} 
to derive the anisotropic LAE-$\alpha$ equations.
However, in \cite{MaSh2002} the second-order Taylor hypothesis
$$
\frac{D}{Dt} \langle \xi'' \rangle \perp u,
$$
where the orthogonality is taken in $L^2$, is necessary to achieve closure.
Our choice of modeling assumptions rendered unnecessary any such
hypothesis on the second-order fluctuations $\xi''$. Second-order Taylor
hypotheses, unlike the first-order hypothesis retained from \cite{MaSh2002},
do not have much precedent in the turbulence literature, as discussed above.
 
\paragraph{Incompressible, Isotropic, Homogeneous Fluids}
To model the motion of an approximately isotropic fluid, we take the covariance
tensor $F$ to be the identity matrix, i.e.
\begin{equation}
\label{eqn:isotropy}
F^{i j} = \Bigl \langle {\xi'}^i {\xi'}^j \Bigr \rangle = \delta^{i j}.
\end{equation}
The choice of $F^{ij} = \delta^{ij}$ is a modeling assumption, and will thus
only be valid for flows which almost preserve this property. Note that
(\ref{eqn:isotropy}) is strictly inconsistent with the advection flow rule, and
thus can only be regarded as an approximation.

For the case of incompressible isotropic mean flow, we assume that
(\ref{eqn:isotropy}) holds; then differentiating this equation with respect to
$x^k$ and $x^j$ and using the fact  that $\xi'$ is divergence-free, we have
$$
\left \langle {\xi'}^i_{,j} {\xi'}^j_{,k} \right \rangle 
= - \left \langle {\xi'}^i_{,jk} {\xi'}^j \right \rangle.
$$
Hence
$$
\left \langle \operatorname{tr} 
(\nabla \xi' \cdot \nabla \xi') \right \rangle =  \left \langle {\xi'}^i_{,j}
{\xi'}^j_{,i} \right \rangle = - \left \langle {\xi'}^i_{,ji} {\xi'}^j\right
\rangle = 0,
$$
and the Lagrangian (\ref{eqn:avg_incomp_aniso_l_expr}) simplifies to
\begin{equation}
\label{eqn:avg_incomp_iso_l_expr}
l^{\alpha}_{\text{in,iso}}(u) = \int_{M} \left \{\frac{1}{2} 
\|u\|^2 + \frac{\alpha^2}{2}\left \langle
\left \| \frac{D \xi'}{Dt} \right \|^2 \right \rangle \right \}dx.
\end{equation}
We emphasize that this is only an approximation, so that
\begin{equation*}
l^{\alpha}_{\text{in,iso}}(u) \approx l^{\alpha}_{\text{in}}(u)
\end{equation*}
along fluid trajectories $u(t)$ for which the covariance tensor is
approximately the identity. Now using the flow rule given by (\ref{incompFL1}), 
the averaged  Lagrangian $l^\alpha_{\text{in}}$ from (\ref{eqn:avg_incomp_iso_l_expr}) becomes
\begin{equation}
\label{eqn:D_norm}
\left \langle \left \| \frac{D \xi'}{Dt} \right \|^2 \right \rangle 
= u^i_{,j} u^i_{,k} \left \langle {\xi'}^j {\xi'}^k \right \rangle = u^i_{,j}
u^i_{,j},
\end{equation}
where we have used the isotropy assumption (\ref{eqn:isotropy}). 
Hence, (\ref{eqn:avg_incomp_iso_l_expr}) becomes
\begin{equation}
\label{eqn:avg_incomp_iso_l_expr_1}
l^{\alpha}_{\text{in}}(u) = \int_{M} \left \{\frac{1}{2} 
\|u\|^2 + \frac{\alpha^2}{2} \left \| \nabla u \right \|^2 \right \}dx.
\end{equation} 
This expression for the averaged Lagrangian 
in the isotropic case is identical to the one derived in \cite{MaSh2001}. 
Now applying either Hamilton's principle or Euler-Poincar\'{e} theory, we
obtain the standard isotropic LAE-$\alpha$ equations:
\begin{subequations}
\label{eqn:LAE_in_iso}
\begin{align}
\label{eqn:LAE_in_iso_1}
\partial_t(1-\alpha^2 \Delta)u 
+ (u\cdot \nabla) (1-\alpha^2 \Delta)u - \alpha^2 (\nabla u )^T \cdot \Delta
u &= -\operatorname{grad} p, \\
\label{eqn:LAE_in_iso_2}
\operatorname{div} u &= 0,
\end{align}
\end{subequations}
where $p$ is the usual fluid pressure.


\section{Averaged Lagrangian for Compressible Flow}
\label{averaging-Euler}
Having understood the incompressible case, we now turn to the compressible
case. The procedure is identical in all aspects except we must now keep track
of density fluctuations.  Start with the reduced Lagrangian for compressible
flow:
\begin{equation}
\label{eqn:unavg_l_expr}
l(u^{\epsilon}, \rho^{\epsilon}) 
= \int_{M} \left( \frac{1}{2} \|u^{\epsilon}\|^2 - W(\rho^{\epsilon}) \right)
\rho^{\epsilon} \ dx.
\end{equation}
The fluid is assumed to be barotropic, meaning that
$W$, the potential energy, is a function only of $\rho$, 
the fluid density.  Now expand the velocity and density in Taylor series
\begin{equation}
\label{eqn:u_and_rho_expan}
\begin{split}
u^{\epsilon} &= u + \epsilon u' 
+ \frac{1}{2} \epsilon^2 u'' + \mathcal{O}(\epsilon^3) \\
\rho^{\epsilon} &= \rho + \epsilon \rho' + \frac{1}{2} \epsilon^2 \rho'' +
\mathcal{O}(\epsilon^3),
\end{split}
\end{equation}
and also expand the potential energy $W$:
$$
W(\rho^{\epsilon}) = W(\rho) 
+ \epsilon W'(\rho) \rho' + \frac{1}{2} \epsilon^2
(W''(\rho) {\rho'}^2 + W'(\rho) \rho'') + \mathcal{O}(\epsilon^3).
$$
Substituting these expansions into the reduced Lagrangian gives
\begin{equation}
\begin{split}
\label{eqn:unavg_expan_l_expr}
l(u^{\epsilon}, \rho^{\epsilon}) &
= \int_{M} \left( \frac{1}{2} \|u\|^2 - W(\rho) \right) \rho \\
 &\, + \epsilon \left[ \left( u \cdot u' - W'(\rho) \rho'\right) \rho 
     + \left( \frac{1}{2} \|u\|^2 - W(\rho) \right) \rho' \right] \\
 &\, + \epsilon^2 \left[ 
       \frac{1}{2} \left( (\|u'\|^2 + u'' \cdot u ) - (W''(\rho) {\rho'}^2
           + W'(\rho) \rho'' ) \right) \rho \right. \\
 &\, \phantom{+\epsilon^2\Biggl[} +\left. ( u \cdot u' - W'(\rho) \rho' ) \rho'
        + \frac{1}{2} \left( \frac{1}{2} \|u\|^2 - W(\rho) \right) \rho''
       \right] + \mathcal{O}(\epsilon^3) \ dx.
\end{split}
\end{equation}
This expansion is now truncated, leaving out all terms of order $\epsilon^3$ and higher.  Denote the truncated Lagrangian by $\hat{l}(u^{\epsilon}, \rho^{\epsilon})$, and define the averaged Lagrangian $l^{\alpha}$ by 
\begin{equation}
\label{eqn:avg_l_def}
l^{\alpha}(u, \rho) = \langle \hat{l}(u^{\epsilon}, \rho^{\epsilon}) \rangle.
\end{equation}
We now outline the procedure by which we arrive at a final written
expression for the averaged Lagrangian $l^\alpha$.  The algebra is 
straightforward but tedious, so details will be omitted.
\begin{enumerate}
\item Use equations (\ref{eqn:u_primes}) and (\ref{eqn:rho_primes})
to rewrite
(\ref{eqn:unavg_expan_l_expr}) in terms of only $u$, $\rho$, and
the fluctuations $\xi'$, $\xi''$.
\item Remove two kinds of terms that vanish inside the average:
\begin{enumerate}
\item linear functions of $\xi'$ or $\xi''$,
\item linear functions of derivatives (either spatial or temporal) of $\xi'$ or $\xi''$.
\end{enumerate}
Note: see ``Zero-Mean Fluctuations'' in \S\ref{averaging-incomp} for
justification.
\item Carry out the averaging operation.  As in the incompressible
case, the only quantities left
inside the average should be nonlinear functions of $\xi'$.
\end{enumerate}
The end result for the averaged Lagrangian for compressible flow is
\begin{equation}
\begin{split}
\label{eqn:avg_comp_l_expr}
l^{\alpha}_{\text{comp}}(u, \rho) &= \int_{M} \left\{ \frac{1}{2} \rho
\|u\|^2 - \rho W(\rho) + \alpha^2 \left[ \frac{1}{2} \rho \left \langle
\left \|\frac{D \xi'}{Dt} \right \|^2 \right \rangle \right. \right. \\
& \left. \left. -\frac{1}{2} w'(\rho) \left \langle \operatorname{div} (\rho \xi')^2 \right \rangle 
- \frac{1}{2} w(\rho) \left \langle \operatorname{div} \operatorname{div} (\rho \xi' \otimes \xi') 
\right \rangle \right] \right\} dx.
\end{split}
\end{equation}
We have introduced $w$, the enthalpy\footnote{Any function $w$
satisfying $\nabla w = (\nabla p)/\rho$, where $p$ is pressure,
is called enthalpy.  Our definition of $w$ implies
$w_{,i} = 2 W'(\rho) \rho_{,i} + \rho W''(\rho) \rho_{,i} = (\rho^2 W'(\rho))_{,i} / \rho = p_{,i} / \rho$ as required.
},
defined by
\begin{equation}
\label{eqn:enthalpy_def}
w(\rho) = W(\rho) + \rho W'(\rho).
\end{equation}


\section{Flow Rule Modeling}
\label{flow-rules}

In deriving the expressions (\ref{eqn:avg_comp_l_expr}) and 
(\ref{eqn:avg_incomp_aniso_l_expr}) for the averaged Lagrangians,  
no assumptions were made regarding how the Lagrangian fluctuations
$\xi'$ evolve. In this section we describe one possible strategy for
modeling $\xi'$.  Note that such a strategy is necessary to achieve
closure for the evolution equations associated with the Lagrangians (\ref{eqn:avg_comp_l_expr}) or (\ref{eqn:avg_incomp_aniso_l_expr}).

\paragraph{Preliminary Observation}
Assuming $\xi'$ evolves via a linear flow rule, as in (\ref{eqn:flow_rule_linear}), 
the vector field $\xi'$ appears in the averaged Lagrangian (\ref{eqn:avg_comp_l_expr})
only as part of the following three expressions\footnote{Similar tensors appear in \cite{Holm1999}; they are referred to as second-order statistics of the Lagrangian fluctuations.}:
\begin{subequations}
\label{eqn:tensors}
\begin{align}
\label{eqn:tensors_1}
F^{ij} &= \left \langle {\xi'}^i {\xi'}^j \right \rangle, \\
\label{eqn:tensors_2}
G^i &= \left \langle {\xi'}^i {\xi'}^j_{,j} \right \rangle, \\
\label{eqn:tensors_3}
H &= \left \langle {\xi'}^i_{,i} {\xi'}^j_{,j} \right \rangle.
\end{align}
\end{subequations}
Note that $F$ is the same \emph{Lagrangian covariance tensor} from the 
incompressible derivation.  In terms of these quantities, the averaged compressible Lagrangian is given in coordinates by
\begin{equation}
\begin{split}
\label{eqn:avg_comp_l_expr_2}
l^{\alpha}_{\text{comp}}(u, \rho) &= \int_{M} \left\{ \frac{1}{2} \rho
u^i u^i - \rho W(\rho) + \alpha^2 \left[ \frac{1}{2} \rho \Omega^{ij} \Omega^{ik} F^{jk} \right. \right. \\
& \left. \left. -\frac{1}{2} w'(\rho) \left(\rho_{,i} \rho_{,j} F^{ij} + 2 \rho \rho_{,j} G^j + \rho^2 H \right)
- \frac{1}{2} w(\rho) \left( \rho F^{ij} \right)_{,ij} \right] \right\} dx.
\end{split}
\end{equation}
Time-differentiating (\ref{eqn:tensors_1}-\ref{eqn:tensors_3}) and using the
linear flow rule (\ref{eqn:flow_rule_linear}) results in
evolution equations for $F$, $G$, and $H$:
\begin{subequations}
\label{eqn:flow_tensors}
\begin{align}
\label{eqn:dF_dt_gen}
\partial_t F^{ij} &= \Omega^{ik} F^{kj} + \Omega^{jk} F^{ki} - u^k F^{ij}_{,k} \\
\label{eqn:dG_dt_gen}
\partial_t G^i &= \Omega^{ik}G^k - u^k G^i_{,k}  + F^{ij} \Omega^{kj}_{,k} + \left \langle {\xi'}^i {\xi'}^j_{,k} \right \rangle (\Omega^{kj}-u^k_{,j}) \\
\label{eqn:dH_dt_gen}
\partial_t H &= 2 \Omega^{ik}_{,i}G^k - u^k H_{,k} + 2 \left \langle {\xi'}^j_{,k} {\xi'}^i_{,i} \right \rangle  (\Omega^{kj}-u^k_{,j}).
\end{align}
\end{subequations}

\paragraph{Flow Rules}
For compressible flows, two flow rules will be considered.  We define them first, 
and then go on to consider their relative merits and demerits:
{
\renewcommand{\theenumi}{\textbf{\Roman{enumi}}}
\renewcommand{\labelenumi}{\theenumi.}
\begin{enumerate}
\item Advection: $\Omega^{ij} = u^i_{,j}$  \label{FL1}
\item Rotation: $\Omega^{ij} = \frac{1}{2} \left( u^i_{,j} - u^j_{,i} \right)$   \label{FL2}
\end{enumerate}
}

\paragraph{Advection}
For our anisotropic model, we shall advect $\xi'$ and treat the quantities $F$, $G$, and $H$ as
parameters in the final system, each of which will have its own evolution equation.  Substituting $\Omega^{ij} = u^i_{,j}$ into the system (\ref{eqn:flow_tensors}) gives
\begin{subequations}
\label{eqn:flow_tensors_FL1}
\begin{align}
\label{eqn:dF_dt_FL1}
\partial_t F &= - \pounds_{u} F \\
\label{eqn:dG_dt_FL1}
\partial_t G &= - \pounds_{u} G + F \cdot \operatorname{grad}(\operatorname{div} u) \\
\label{eqn:dH_dt_FL1}
\partial_t H &= 2 \operatorname{grad}(\operatorname{div} u) \cdot G - u \cdot \operatorname{grad} H.
\end{align}
\end{subequations}
One advantage of the advection flow rule is that it automatically closes the
system (\ref{eqn:flow_tensors}).  For a general choice of $\Omega$, the system
involves $\left \langle {\xi'}^i {\xi'}^{j}_{,k} \right \rangle$ and
$\left \langle {\xi'}^{j}_{,k} {\xi'}^{i}_{,i} \right \rangle$, which cannot be expressed
solely in terms of $F$, $G$, and $H$.

\paragraph{Rotation}
For our isotropic model, we want to know whether the evolution equation (\ref{eqn:dF_dt_gen}) for $F$ preserves the isotropy relationship $F = \text{Identity}$.  Suppose $F^{ij} = \delta^{ij}$ at $t = 0$.  Then substituting into (\ref{eqn:dF_dt_gen})
reveals that
\begin{equation}
\label{eqn:dF_dt_isotropy}
\partial_t |_{t = 0} F^{ij} = \Omega^{ij} + \Omega^{ji}.
\end{equation}
If $\Omega$ is antisymmetric, we have $\partial_t |_{t = 0} F = 0$, and $F(x,t) = \text{Identity}$ solves (\ref{eqn:dF_dt_gen}) for all $t$.  We wish to know whether this solution is unique.  This is guaranteed by a straightforward generalization of the results concerning linear hyperbolic systems of first-order equations from \cite{Evans1998}, assuming sufficient smoothness of $u$.

We conclude that
antisymmetry of $\Omega$ is sufficient to guarantee that the initial data $F = \text{Identity}$ is in fact preserved for all $t$.  
Then an immediate choice of a tensor $\Omega$ that
is antisymmetric is given by the rotation flow rule (\ref{FL2}).
This form has a very attractive physical interpretation. Putting the linear flow rule equation
(\ref{eqn:flow_rule_linear}) together with (\ref{FL2}) gives us
\begin{equation}
\label{eqn:flow_rule_rotation}
\frac{D \xi'}{Dt} = \omega \times \xi',
\end{equation}
where $\omega = \operatorname{curl} u$ is the vorticity vector. The last equation 
can be interpreted in the sense that fluctuations are rigidly transported by the 
mean flow, with a local angular velocity given by the vorticity vector.

Finally, the rotation flow rule (\ref{FL2}) does not by itself close the system (\ref{eqn:flow_tensors}).  When using this flow rule, we shall assume that $G = 0$ and
$H = \beta^2$.  


\section{Equations for Averaged Dynamics}
\label{averaged-pdes}
Here we shall write down two systems of coupled PDEs which describe the
evolution of the average velocity and density in a compressible flow.  Each PDE
is derived from an associated averaged Lagrangian.
\paragraph{Compressible, fully Anisotropic, Inhomogeneous Fluids}
By substituting (\ref{FL1}) into the Lagrangian (\ref{eqn:avg_comp_l_expr}), we obtain closure: the 
Lagrangian no longer depends explicitly on $\xi'$, but instead on the tensors $F$, $G$, and $H$, 
for which a self-contained system of evolution
equations (\ref{eqn:flow_tensors_FL1}) has already been derived---see \S\ref{flow-rules} for details. 
Applying  Hamilton's principle directly to (\ref{eqn:avg_comp_l_expr_2}) yields an evolution equation for $u$, the average fluid velocity.  We write
this equation  using the operator $\mathcal{A}$, 
which is defined as
\begin{equation}
\label{eqn:A_operator_def}
\left( \mathcal{A} v \right)^i = \frac{1}{\rho}
\left( \rho v^i_{,j} F^{jk} \right)_{,k}.
\end{equation}
We also write $\tilde{w} = \rho w'(\rho)$ where $'$ means $d / d\rho$ as usual.
The anisotropic compressible LAE-$\alpha$ equations are:
\begin{subequations}
\label{eqn:avg_advection_pde_short}
\begin{multline}
\left(\partial_t {u}^n + u^n_{,i} u^i \right)
= (1 - \alpha^2 \mathcal{A})^{-1} 
\frac{1}{\rho} \Biggl\{ - \rho w_{,n} - \frac{\alpha^2}{2} \left[ 
\rho \left( F^{ij} u^k_{,i} u^k_{,j} \right)_{,n}
+ F^{ij}_{,ij} \rho \tilde{w}_{,n} \Biggr. \right. \\
+ F^{ij}_{,n} \rho_{,i} \tilde{w}_{,j} 
+ \left( F^{ij}_{,n} \rho \right)_{,ij} \tilde{w} 
+2 \Biggl. \left. G^i_{,n} \rho \tilde{w}_{,i} 
+ 2 G^i \left( \rho \tilde{w}_{,n} \right)_{,i} 
+ \left( H \rho^2 \tilde{w}' \right)_{,n} \right]
\Biggr\}
\end{multline}
\vspace{-0.75cm}
\begin{align}
\partial_t {\rho} &
= - \operatorname{div} (\rho u) \\
\partial_t {F} &
= - \nabla F \cdot u + F \cdot \nabla u + \nabla u^T \cdot F \\
\partial_t {G} &
= - u \cdot \nabla G + G \cdot \nabla u + F \cdot 
\operatorname{grad}(\operatorname{div} u) \\
\partial_t {H} &
= 2 \operatorname{grad}(\operatorname{div} u) \cdot G - u \cdot
\operatorname{grad} H.
\end{align}
\end{subequations}

\paragraph{Well-posedness}
We now sketch a rough well-posedness argument for the system
(\ref{eqn:avg_advection_pde_short}).  Assume that the tensor $F$ is
positive-definite.  By this it is meant, since $F$ is  a $(2,0)$ tensor, that for
any one-form $\theta$, the contraction
$F : (\theta \otimes \theta)$ is positive everywhere.
Given the
$\rho$-weighted inner product $\langle f, g \rangle = \int f \, g \, \rho$,
we have $\langle f, -\mathcal{A} f \rangle = -\int f \, 
\frac{1}{\rho} \left( \rho f_{,j} F^{jk} \right)_{,k} \, \rho 
 = \int f_{,j} F^{jk} f_{,k} \, \rho > 0$.
Since $-\mathcal{A}$ is a positive definite linear operator, 
$(1 - \alpha^2 \mathcal{A})$ 
has trivial kernel and we expect that (\ref{eqn:avg_advection_pde_short}) is
well-posed.

It would be of analytical interest to see to what extent the ``geodesic
part'' of these equations define a smooth spray in the sense of
\cite{EbMa1970}, and which holds for the EPDiff equations (that is, the
$n$-dimensional CH equations), as explained in \cite{HoMa2004}.

\paragraph{Compressible, isotropic, inhomogeneous}
For this case we use flow rule (\ref{FL2}), which can be written in vector
notation as 
$$
\Omega = \frac{1}{2} \left( \nabla u - \nabla u^T \right).
$$
Recall that this flow
rule is compatible with an isotropic choice of the covariance tensor,
i.e. $F^{ij} = \delta^{ij}$.  We further assume
that $G = 0$ and $H = \beta^2$ for some constant $\beta$.
Using flow rule (\ref{FL2}) along with these extra assumptions in the general
Lagrangian expression (\ref{eqn:avg_comp_l_expr_2}) gives us a Lagrangian in only two variables:
\begin{multline}
\label{eqn:avg_comp_l_isotropic}
l(u, \rho) = \int_M \left( \frac{1}{2} \rho \| u \|^2 - \rho W(\rho)
+ \alpha^2 \left[ \frac{1}{4} \rho \left( \| \nabla u \|^2 
- \operatorname{tr} \left(\nabla u \cdot \nabla u \right) \right) \right. \right. \\
- \left. \left. \frac{1}{2} w'(\rho) \left( \| \nabla \rho \|^2 + \rho^2 \beta^2 \right) - \frac{1}{2} w(\rho) \Delta \rho \right] \right) \ d^N \! x,
\end{multline}
where $w(\rho) = W(\rho) + \rho W'(\rho)$ is the enthalpy introduced in 
(\ref{eqn:enthalpy_def}).  Regarding this as a Lagrangian in $u$ and $\mu = \rho \, d^N \! x$,
one uses the semidirect product Euler-Poincar\'e equations (see \cite{HoMaRa1998b}) to derive the system
\begin{subequations}
\label{eqn:avg_comp_pde_isotropic}
\begin{align}
\partial_t (\rho v) + \left( u \cdot \nabla \right) (\rho v) 
+ \alpha^2 \operatorname{div}
 \left( \rho \Omega \cdot \nabla u \right) + \rho v \operatorname{div} u
 &= - \nabla \tilde{p} \\
 \partial_t \rho + \operatorname{div} (\rho u) &= 0.
\end{align}
\end{subequations}
with the modified momentum $\rho v$ and modified pressure $\tilde{p}$ given by
\begin{align}
\label{eqn:v_def}
\rho v &= \rho u + \alpha^2 \operatorname{div} \left( \rho \Omega \right), \\
\label{eqn:ptilde_def}
\nabla \tilde{p} &= \nabla p + \alpha^2 \beta^2 \rho \, \nabla \left( \rho w' + \frac{1}{2} \rho^2 w'' \right).
\end{align}
Here are explicit coordinate expressions for two slightly complicated objects:
\begin{align*}
\rho v^i &=
 \rho u^i + \frac{1}{2} \alpha^2 \left( \rho \left( u^j_{,i} - u^i_{,j} \right) \right)_{,j} \\
 \operatorname{div} \left( \rho \Omega \cdot \nabla u \right) &=
\left( \rho \Omega^{ki} u^{i}_{,j} \right)_{,k}.
\end{align*}
The following convention for divergences of tensors has been used:
given a 2-tensor $A^{ij}$, we set
$$
(\operatorname{div} A)^j = A^{ij}_{,i}.
$$
That is, the contraction implicit in the divergence 
operation always takes place on the \emph{first} index.  

\paragraph{Observations}
\begin{itemize}
\item In the case of homogeneous incompressible flow, where $\rho$ is
constant and
$\operatorname{div} u = 0$, the definition of $\rho v$ in (\ref{eqn:v_def})
reduces to
$$
v = \left(1 - \frac{1}{2} \alpha^2 \Delta \right) u,
$$
which after rescaling $\alpha$ to get rid of the factor of $1/2$ is precisely
the $v$ one finds in treatments of the incompressible LAE-$\alpha$ and
LANS-$\alpha$ equations.
\item The above does not work in one spatial dimension.  The problem is that here $\Omega$ reduces to $(u_x - u_x)/2 = 0$, which clearly does not describe transport at
all. For a 1-D isotropic model one may very well want to forget about
antisymmetry of $\Omega$ and instead use something such as the advection flow
rule.  One may, quite reasonably, conclude that the only meaning of isotropy in
1-D should be reflection symmetry.
\end{itemize}


\section{Future Directions}
\label{future-directions}

\paragraph{The Initialization Problem}
Perhaps the largest unsolved problem for the Lagrangian averaged equations is
the initialization problem.  A concise statement of the problem reads:
\begin{quote}
Given initial data $u_0(x)$ for the Euler equation, how does one 
obtain initial data $U_0(x)$ for the LAE-$\alpha$ equation?
\end{quote}
Let us look at this problem in slightly more detail. Let $u$ denote the
solution of  the incompressible Euler equations for initial data $u_0$, i.e.
$u(x,0) = u_0(x)$. Similarly, let $U$ denote the solution of the
incompressible, isotropic LAE-$\alpha$  equations (\ref{eqn:LAE_in_iso}) for
initial data $U_0$.  

Now $U$ should be, in some sense, the mean flow of the fluid.  This means
that $U_0$ should be the mean flow of the fluid at time $t = 0$, implying
that $U_0$ should be, in some sense, an ``averaged'' or ``filtered'' version of
$u_0$.  The question is: how does one derive $U_0$ from $u_0$?  Another way
of phrasing this question is: how do we describe (approximately) the initial state of the
fluid (given exactly, for our purposes, by the field $u_0$) using only
the mean flow variable $U_0$?

Numerous methods have been used to initialize the LAE-$\alpha$ equations for
use in numerical simulations, but none of these methods has any theoretical
foundation.   There is also no theory regarding how one should filter a full
Euler flow $u$, or even a family of flows $u^\epsilon$, in order to obtain a
mean flow that could be compared with the full LAE-$\alpha$ trajectory $U$. 
In this respect, equation (\ref{eqn:u-centered}), which states that
$$
\langle u^\epsilon \circ \xi^\epsilon (x,t) \rangle = u(x,t),
$$
is not helpful: we have no way to compute the fluctuation diffeomorphism
group $\xi^\epsilon$.  Therefore we have no way to compute the left-hand side 
$\langle u^\epsilon \circ \xi^\epsilon \rangle$.

The difficulty can be summarized in the following commutative diagram.  Here
$S$ is the standard Euler action and $S^\alpha$ is the Lagrangian-averaged
action.
\begin{equation*}
\qquad \qquad \qquad \xymatrix@C=1.1in{
S \ar^{\txt{\footnotesize Lagrangian average}}[r] \ar[d] & S^{\alpha}
\ar^{\txt{\footnotesize derive PDE, \\ \footnotesize solve numerically}}[d] \\
u \ar@{-->}_{\txt{\footnotesize the missing link}}[r] & U
}
\end{equation*}
Solid arrows represent steps that we know how to carry out.  The dashed arrow
represents  the one step that we do not know how to carry out.  Our strategy
for this problem will be  to develop methods by which we can test different
filters for obtaining $U_0$ from $u_0$  in practice.  

\paragraph{Treatment of Densities}
Another area for further investigation involves our treatment of the density
tube $\mu^\epsilon$.  There are two questions to ground us:
\begin{enumerate}
\item We have tacitly assumed that at $t = 0$, and for all
$\epsilon$, all $x$,
$$
\mu(x,t) = \mu^\epsilon(x,t).
$$
An argument similar to the one made above in our discussion of the
initialization problem can be made here.  Namely, $\mu(x,0)$ represents the
mean density at time
$t = 0$.  Meanwhile, $\mu^\epsilon(x,0)$ represents the true density of the
fluid at time $t = 0$.  These two quantities need not be equal.  This prompts
the question:  how would we carry out the procedure from Sections
\ref{averaging-general} and 
\ref{averaging-Euler} with tubes in which each trajectory does \emph{not} have
the same initial density $\mu(x,0)$?
\item As our derivation of the averaged compressible equations stand, we have
derived the fact that the ``mean'' density $\mu$ was advected by the mean
flow $U$:
$\partial_t \mu = - \pounds_{U} \mu$. Substituting $\mu = \rho \, d^N \! x$
and using the definition of  divergence yields the standard continuity
equation
$$\partial_t \rho + \operatorname{div} (\rho U) = 0.$$
In both RANS and LES treatments of averaged/filtered flow, the mean flow $U$
satisfies a \emph{modified} continuity equation rather than the standard
one.  Therefore: why does the Lagrangian averaged mean density $\mu$ satisfy
the  usual continuity equation?
\end{enumerate}
The two questions regarding densities are in fact related.  To see this, let
us suppose  that given the initial density $\mu_0$ associated with the center
line of our tube $\eta$, we have a method for constructing a family of
initial densities $\mu^\epsilon_0$ for each of the other curves in the tube
$\eta^{\epsilon}$.  Now defining\footnote{Note that $\langle
\mu^\epsilon(x,t) \rangle \neq \mu(x,t).$}
$$
\mu^\epsilon (t) = (\eta^\epsilon_t)_{\ast} \mu^\epsilon_0 \ \ \text{ and } \ \ \bar{\mu} (t) = \langle \mu^\epsilon(t) \rangle,
$$
we will find that $\bar{\mu} (t) = \bar{\rho} (t) \, d^N \! x$ satisfies a modified continuity equation
$$
\partial_t \bar{\rho}(t) + \operatorname{div}(\bar{\rho} u) + 
\operatorname{div} \left \langle \rho^\epsilon \left( \epsilon u' 
+ \frac{1}{2} \epsilon^2 u'' \right) \right \rangle = 0.
$$
To close this equation, we must either carry out the average directly, or we must expand
$\rho^\epsilon$ about a suitable trajectory and make modeling assumptions. 

\paragraph{Filtered Lagrangians}
We have seen that the current averaging procedure leads to complicated
averaged equations. Furthermore, there is no clear way to evaluate numerically the
flow rules we have proposed on  physical grounds.  One of our immediate goals
is to investigate a filtering approach, still at the level of the Lagrangian,
which will lead to simpler averaged models that can be tested numerically. 
The filtering approach we have in mind begins with a decomposition of the
velocity field
\begin{equation}
\label{eqn:u_and_rho_decomp}
u = \bar{u} + u' \ \text{ and } \ \rho = \bar{\rho} + \rho'
\end{equation}
into mean and fluctuating components.  This would replace the Taylor expansion 
(\ref{eqn:u_and_rho_expan}) of $u^\epsilon$ and $\rho^\epsilon$ that we carried out in the
present work, and would therefore lead to Lagrangians and equations with much
less algebraic complexity.  
As opposed to the axiomatic averaging operation $\langle \cdot \rangle$,
the filter shall be specified concretely.  We expect this to help greatly
with the initialization and density problems discussed above; furthermore,
the filtering approach leads naturally to questions about the relationship
between LES and LAE-$\alpha$ models.

\paragraph{Simpler Models}
As we previously noted, the flow rule approach developed in this paper does not yield a one-dimensional compressible averaged model.  We are currently investigating such a model, derived from the filtered Lagrangian
\begin{equation}
\label{eqn:one_d_filter}
l(\rho,u) = \int \biggl( \frac{1}{2} u v - W(\rho) \biggr) \, \rho \, d^N \! x,
\end{equation}
where $v = \bigl(1 - \alpha^2 \partial_{xx} \bigr) u$.  To derive this Lagrangian, we filter only the velocity, leaving density and potential energy alone.  This is the compressible analogue of the filtered Lagrangian used in deriving the Camassa-Holm equation \cite{CaHo1993}.  The analysis and numerical simulation of the new equations presented in Section \ref{averaged-pdes} of this work will be difficult.  Much easier is the analysis of the PDE associated with (\ref{eqn:one_d_filter}).  In particular, we expect that numerical studies of this one-dimensional model will yield insight into the dynamics of the higher-dimensional equations.

\paragraph{Entropy}
In the derivation of our compressible averaged models, we have made the barotropic assumption $W = W(\rho)$.  We expect the resulting barotropic model to be useful in computing mean flow quantities in regimes where we are not concerned with strong physical shocks, for example in climate models.  The next major step forward will be to remove the barotropic assumption, and derive a model that is valid in regimes where we are concerned with shocks.

To this end, we have derived an averaged model for the general case, where the potential energy has the form $W(\rho,S)$, where $S$ is the entropy.  This model, which consists of a system of equations for $\rho$, $u$, and $S$, also involves the pressure $p$.  Therefore, in order to close the system, we require an equation of state relating $p$ to $\rho$ and $S$.  The open question now is as follows: given an equation of state for the compressible Euler system, what is the equation of state relating the averaged variables to one another?  In other words, how does Lagrangian averaging interact with the thermodynamics of the system?  We hope that analyzing a finite-dimensional case of this interaction will shed light on this issue.

\paragraph{Connections with Kevrekidis' Coarse/Fine Methods}
Given a description of any mechanical system, not necessarily involving fluids, in the form 
of a Lagrangian $\ell$, we can carry out the procedure described in 
\S\ref{averaging-general} to find an averaged Lagrangian $\langle \ell
\rangle$. From this we can derive equations of motion for the average
dynamics of the original system. Changing our language slightly, we say that
we have a general method for extracting the ``coarse'' dynamics of a
mechanical system whose full description involves motions on both fine and
coarse scales.

Another method for computing the coarse-scale dynamics of a mechanical system has been 
put forth by in \cite{Ke2003}.  Kevrekidis' method does not
involve trying to write down equations of motion which govern the coarse dynamics.  
Instead, he offers an algorithmic approach, the crux of which is as follows.  The coarse dynamics of a system are found by \emph{lifting} the initial ($t = t_0$) state to an 
ensemble of initial states, \emph{integrating} each using the full equations until some small
final time $t = \epsilon$ has been reached, and \emph{projecting} the 
resulting $t = \epsilon$ states onto a single state. This $t = \epsilon$ state is then \emph{extrapolated} to a state at some desired $t = t_f > 0$.  By iterating this process and
tuning the lifting, projection, and extrapolation operations, this method can be used to
recover the coarse dynamics of the system.

Now the question that begs to be asked is as follows: for the case of fluid dynamics, how
different are the coarse dynamics provided by the LANS-$\alpha$ equation from the coarse
dynamics one would obtain by following Kevrekidis?  The difficulty in answering this
question lies in implementing a full fine-scale integrator for fluids that one could 
successfully embed inside Kevrekidis' coarse-scale algorithm.  We look forward to tackling
this task soon.

\section{Acknowledgments}
\label{acks}
We extend our sincerest thanks to Steve Shkoller, Darryl Holm and Marcel
Oliver for helpful discussions and  criticism on a wide array of issues
central to this paper. The research in this paper was partially
supported by AFOSR Contract F49620-02-1-0176. Harish S. Bhat thanks the
National Science Foundation for supporting him with a Graduate Research
Fellowship.


\section{Appendix: Fluctuation Calculus Details}
\label{appendix}
Before proceeding with any derivations, we state the Lie derivative theorem for
time-dependent vector fields: if the vector field $X_{\lambda}$ has flow $F_{\lambda}$, then
\begin{equation}
\label{eqn:lie_deriv_thm}
\frac{d}{d \lambda} F^{\ast}_{\lambda} Y_{\lambda} = F^{\ast}_{\lambda} \left(
\frac{\partial Y_{\lambda}}{\partial \lambda} + \pounds_{X_{\lambda}} Y_{\lambda} \right).
\end{equation}

Our task now is to derive equations (\ref{eqn:rho_primes}).  
Starting with (\ref{eqn:rho_mod}), let us move $\eta^\epsilon$ to the right-hand side of the equation:
\begin{equation}
\label{eqn:rho_mod_pb}
\mu_0 = \left( \eta^\epsilon \right)^{\ast} \mu^\epsilon.
\end{equation}
The strategy is to differentiate with
respect to $\epsilon$ and use the Lie derivative theorem (\ref{eqn:lie_deriv_thm}).
The intrinsic definition of divergence
\begin{equation}
\label{eqn:div_def}
\pounds_{\zeta} (\nu) = \left( \operatorname{div}_{\nu} \zeta \right) \nu
\end{equation}
and the canonical volume form $d^N \! x = dx^1 \wedge \cdots \wedge dx^N$ will both be used in what follows.
Note that $\operatorname{div} \zeta$ with no subscript on the $\operatorname{div}$
means $\pounds_{\zeta} (d^N \! x)$.  Before applying the Lie derivative theorem, note
that the vector field
\begin{equation}
\label{eqn:w_eps}
W^{\epsilon} = \frac{\partial}{\partial \epsilon} \eta^{\epsilon} \circ
\left( \eta^{\epsilon} \right)^{-1}
\end{equation}
has flow $\eta^\epsilon$.  A simple computation yields
\begin{equation}
\label{eqn:W_prime}
\left. \frac{\partial}{\partial \epsilon} \right|_{\epsilon = 0} W^{\epsilon} = \xi'' - \nabla \xi' \cdot \xi'. 
\end{equation}
Then we start with $\rho'$:
\begin{alignat*}{2}
\frac{\partial}{\partial \epsilon} \mu_0 = 0 &= \frac{\partial}{\partial \epsilon}
\left( \eta^\epsilon \right)^{\ast} \mu^\epsilon & \qquad & \text{by differentiating (\ref{eqn:rho_mod_pb})} \\
 &= \left( \eta^\epsilon \right)^{\ast} \left( \frac{\partial \mu^\epsilon}{\partial \epsilon} + \pounds_{W^\epsilon} \mu^\epsilon \right) & \qquad & \text{by (\ref{eqn:lie_deriv_thm})} \\
 &= \eta^{\ast} \left( \mu' + \pounds_{\xi'} \mu \right) & \qquad & \text{at } \epsilon = 0 \\
\Longrightarrow \mu' &= - \pounds_{\xi'} \mu \\
\rho' \, d^N \! x &= - \left( \pounds_{\xi'} \rho \right) d^N \! x
                  - \rho \left( \pounds_{\xi'} d^N \! x \right) & \qquad &
                    \text{by (\ref{eqn:mu_defs})} \\
\rho' \, d^N \! x &= - \left( \nabla \rho \cdot \xi' + \rho \operatorname{div} \xi'
                    \right) d^N \! x & \qquad & \text{by (\ref{eqn:div_def})} \\
\Longrightarrow \rho' &= - \operatorname{div} \left( \rho \xi' \right)
\end{alignat*}
Next we compute $\rho''$:
\begin{align*}
\frac{\partial^2}{\partial \epsilon^2} \mu_0 = 0 &= \frac{\partial^2}{\partial \epsilon^2}
\left( \eta^\epsilon \right)^{\ast} \mu^\epsilon \\
&= \left( \eta^{\epsilon} \right)^{\ast} \left( 
\frac{\partial^2}{\partial \epsilon^2} \mu^{\epsilon} + \pounds_{W^\epsilon}
\frac{\partial \mu^\epsilon}{\partial \epsilon} +
\frac{\partial}{\partial \epsilon} \left( \pounds_{W^\epsilon} \mu^\epsilon
\right) + \pounds_{W^\epsilon} \pounds_{W^\epsilon} \mu^\epsilon \right) \\
\Longrightarrow 0 &= \eta^{\ast} \left( \mu'' + 2 \pounds_{\xi'} \mu'
+\pounds_{\xi'' - \nabla \xi' \cdot \xi'} \mu 
+ \pounds_{\xi'} \pounds_{\xi'} \mu \right) \\
\Longrightarrow \mu'' 
&= - \pounds_{\xi''} \mu + 2 \pounds_{\xi'} \pounds_{\xi'} \mu
- \pounds_{\xi'} \pounds_{\xi'} \mu 
                    + \pounds_{\nabla \xi' \cdot \xi'} \mu \\
\rho'' \, d^N \! x &= - \left( \operatorname{div} \left( \rho \xi'' \right) \right) d^N \! x
+ \operatorname{div} \left( \operatorname{div} \left( \rho \xi' \right) \xi' \right) d^N \! x
+ \operatorname{div} \left( \rho \nabla \xi' \cdot \xi' \right) d^N \! x \\
\Longrightarrow \rho'' &= - \left( \operatorname{div} \left( \rho \xi'' \right) \right)  
+ \left( \left( \rho {\xi'}^i \right)_{,i} {\xi'}^j \right)_{,j}
+ \left( \rho {\xi'}^j_{,i} {\xi'}^i \right)_{,j} \\
&= - \left( \operatorname{div} \left( \rho \xi'' \right) \right)  
+ \left( \rho {\xi'}^i {\xi'}^j \right)_{,ij} \\
&= - \left( \operatorname{div} \left( \rho \xi'' \right) \right)  
+ \operatorname{div} \operatorname{div} \left( \rho \xi' \otimes \xi' \right)
\end{align*}




\begin{thebibliography}{99}



\bibitem[Abraham, Marsden, and Ratiu(1988)]{AbMaRa1988}
Abraham, R., J.~E. Marsden, and T.~S. Ratiu [1988], {\em Manifolds, Tensor
  Analysis and Applications}, volume~75 of {\em Applied Mathematical Sciences}.
\newblock Springer-Verlag, New York, second edition.

\bibitem[Andrews and McIntyre(1978)]{AnMc1978}
Andrews, D. and M.~E. McIntyre [1978], An exact theory of nonlinear waves on a
  Lagrangian-mean flow, {\em J. of Fluid Mech.} \textbf{89}, 609--646.


\bibitem[Camassa and Holm(1993)]{CaHo1993}
Camassa, R. and D.~D. Holm [1993], An integrable shallow water equation with peaked solitons, {\em Phys. Rev. Lett.} \textbf{71}, 1661--1664.

\bibitem[Chen et~al.(1998)Chen, Foias, Holm, Olson, Titi, and
  Wynne]{ChFoHoOlTiWy1998}
Chen, S.~Y., C.~Foias, D.~D. Holm, E.~J. Olson, E.~S. Titi, and S.~Wynne
  [1998], The {C}amassa--{H}olm equations as a closure model for turbulent
  channel and pipe flow, {\em Phys. Rev. Lett.} \textbf{81}, 5338--5341.


\bibitem[Chen et~al.(1999)Chen, Holm, Margolin, and Zhang]{ChHoMaZh1999}
Chen, S.~Y., D.~D. Holm, L.~G. Margolin, and R.~Zhang [1999], Direct numerical
  simulations of the {N}avier-{S}tokes alpha model, {\em Physica D}
  \textbf{133}, 66--83.


\bibitem[Cocke(1969)]{Cocke1969}
Cocke, W.~J. [1969], Turbulent Hydrodynamic Line Stretching: Consequences of
  Isotropy, {\em Phys. Fluids} \textbf{12}, 2488--2492.



\bibitem[Ebin and Marsden(1970)]{EbMa1970}
Ebin, D.~G. and J.~E. Marsden [1970], Groups of diffeomorphisms and the motion
  of an incompressible fluid, {\em Ann. of Math.} \textbf{92}, 102--163.

\bibitem[Evans(1998)]{Evans1998}
Evans, L.~E. [1998], {\em Partial Differential Equations}, volume~19 of {\em Graduate Studies in Mathematics}.  \newblock American Mathematical Society, Providence, RI.



\bibitem[Foias, Holm, and Titi(2002)]{FoHoTi2002}
Foias, C., D.~D. Holm, and E.~S. Titi [2002], The three dimensional viscous
  Camassa-Holm equations and their relation to the Navier-Stokes equations and
  turbulence theory, {\em Dyn. and Diff. Eqns.} \textbf{14},
  1--36.


\bibitem[Holm(1999)]{Holm1999}
Holm, D.~D. [1999], Fluctuation effects on 3D Lagrangian mean and Eulerian mean
  fluid motion, {\em Physica D} \textbf{133}, 215--269.


\bibitem[Holm(2002a)]{Holm2002}
Holm, D.~D. [2002a], Averaged Lagrangians and the mean effects of
fluctuations in
  ideal fluid dynamics, {\em Physica D} \textbf{170}, 253--286.

\bibitem[Holm(2002b)]{Holm2002b}
Holm, D.~D. [2002b], Lagrangian averages, averaged {L}agrangians, and the mean
  effects of fluctuations in fluid dynamics, {\em Chaos} \textbf{12}, 518--530.

\bibitem[Holm and Marsden(2004)]{HoMa2004}
Holm, D. and J.~E. Marsden [2004], Peakons, Filaments and Sheets for the
  EPDiff Equation, {\em Festschrift for Alan Weinstein, Birkhauser, Boston (to
  appear)}.

\bibitem[Holm, Marsden, and Ratiu(1998a)]{HoMaRa1998a}
Holm, D.~D., J.~E. Marsden, and T.~S. Ratiu [1998a], {E}uler--{P}oincar\'e
  models of ideal fluids with nonlinear dispersion, {\em Phys. Rev. Lett.}
  \textbf{349}, 4173--4177.

\bibitem[Holm, Marsden, and Ratiu(1998b)]{HoMaRa1998b}
Holm, D.~D., J.~E. Marsden, and T.~S. Ratiu [1998b], The
{E}uler--{P}oincar\'{e} equations and semidirect products with applications
to continuum theories, {\em Adv. in Math.} \textbf{137}, 1--81.



\bibitem[Kawahara(1970)]{Kawahara1970}
Kawahara, T. [1970], Weak nonlinear magneto-acoustic waves in a cold plasma
in the presence of effective electron-ion collisions, {\em J. Phys. Soc.
Japan} \textbf{28}, 1321--1329.


\bibitem[Kakutani and Kawahara(1970)]{KaKa1970}
Kakutani, T. and T.~Kawahara [1970], Weak ion-acoustic shock waves, {\em J.
  Phys. Soc. Japan} \textbf{29}, 1068--1073.


\bibitem[Kevrekidis et. al.(2003)]{Ke2003}
Kevrekidis, I.~G., C.~W. Gear, J.~M. Hyman, P.~G. Kevrekidis, O. Runborg, C. Theodoropoulos [2002],  Equation-free, coarse-grained multiscale computation: enabling microscopic simulators to perform system-level analysis, {\em Communications in the Mathematical Sciences}
\textbf{1}, 715--762.



\bibitem[Lax and Levermore(1983)]{LaLe1983}
Lax, P.~D. and C.~D. Levermore [1983], The small dispersion limit of the KdV
  equations. III, {\em Comm. Pure Appl. Math.} \textbf{XXXVI}, 809--830.


\bibitem[Liepmann and Roshko(1957)]{LiRo1957}
Liepmann, H.~W. and A.~Roshko [1957], {\em Elements of Gasdynamics}.
\newblock Wiley, New York.



\bibitem[Marsden, Ratiu, and Weinstein(1984)]{MaRaWe1984}
Marsden, J.~E., T.~S. Ratiu, and A.~Weinstein [1984], Semi-direct products and
  reduction in mechanics, {\em Trans. Amer. Math. Soc.} \textbf{281}, 147--177.

\bibitem[Marsden and Shkoller(2001)]{MaSh2001}
Marsden, J.~E. and S.~Shkoller [2001], Global well-posedness of the
  LANS-$\alpha$ equations, {\em Proc. Roy. Soc. London} \textbf{359},
  1449--1468.

\bibitem[Marsden and Shkoller(2003)]{MaSh2002}
Marsden, J.~E. and S.~Shkoller [2003], The anisotropic averaged Euler
  equations, {\em Arch. Rat. Mech. An.} \textbf{166}, 27--46.


\bibitem[Mohseni et~al.(2003)Mohseni, Kosovi\'{c}, Shkoller, and
  Marsden]{MoKoShMa2003}
Mohseni, K., B.~Kosovi\'{c}, S.~Shkoller, and J.~E. Marsden [2003], Numerical
  simulations of the Lagrangian averaged Navier-Stokes equations for
  homogeneous isotropic turbulence, {\em Physics of Fluids} \textbf{15},
  524--544.



\bibitem[Rivlin and Erickson(1955)]{RiEr1955}
Rivlin, R. and J.~L. Erickson [1955], Stress-deformation relations for
  isotropic materials, {\em J. Rat. Mech. Anal.} \textbf{4}, 323--425.



\bibitem[Shapiro(1953)]{Shapiro1953}
Shapiro, A.~H. [1953], {\em The dynamics and thermodynamics of compressible
  fluid flow}.
\newblock Ronald Press Co., New York.



\bibitem[Taylor(1938)]{Taylor1938}
Taylor, G.~I. [1938], The spectrum of turbulence, {\em Proc. R. Soc. London
  Ser. A} \textbf{164}, 476--490.



\bibitem[Whitham(1974)]{Whitham1974}
Whitham [1974], {\em Linear and Nonlinear Waves}.
\newblock Wiley-Interscience.

\end{thebibliography}
\end{document}